\documentclass[sigconf,letterpaper,nonacm]{acmart}

\usepackage{multirow}
\usepackage{here}

\AtBeginDocument{%
  \providecommand\BibTeX{{%
    \normalfont B\kern-0.5em{\scshape i\kern-0.25em b}\kern-0.8em\TeX}}}

\copyrightyear{2021}
\acmYear{2021}

\setcopyright{othergov}
\acmConference[AINTEC '21]{Asian Internet Engineering Conference}{December 14--16, 2021}{Virtual Event, Japan}
\acmBooktitle{Asian Internet Engineering Conference (AINTEC '21), December 14--16, 2021, Virtual Event, Japan}
\acmPrice{15.00}
\acmDOI{10.1145/3497777.3498547}
\acmISBN{978-1-4503-9184-9/21/12}

\begin{document}

\title{MAC address randomization tolerant crowd monitoring system using Wi-Fi packets}

\author{Yuyi Cai}
\affiliation{%
	\institution{The University of Tokyo}
	\state{Tokyo}
	\country{Japan}}
\email{cai@hongo.wide.ad.jp}

\author{Manabu Tsukada}
\affiliation{%
  \institution{The University of Tokyo}
  \country{Japan}}
\email{mtsukada@g.ecc.u-tokyo.ac.jp}

\author{Hideya Ochiai}
\affiliation{%
  \institution{The University of Tokyo}
  \country{Japan}}
\email{jo2lxq@hongo.wide.ad.jp}

\author{Hiroshi Esaki}
\affiliation{%
  \institution{The University of Tokyo}
  \country{Japan}}
\email{hiroshi@wide.ad.jp}

\renewcommand{\shortauthors}{Cai}

\begin{abstract}
% Due to the rapid spread of wireless devices, researchers are trying to analyze people’s position and action by using MAC addresses inside probe requests, one kind of Wi-Fi packet. 
% These data could be used for crowdedness estimation, marketing, hazard maps, and other beneficial activities. They thought MAC addresses were unique, and we could count the number of devices by counting the number of MAC addresses. However, from 2014, vendors began putting MAC address randomization systems into their devices to protect user’s privacy. When this system is turned on, devices change their MAC addresses once every few minutes. 
% 調整あり
Media access control (MAC) addresses inside Wi-Fi packets can be used for beneficial activities such as crowdedness estimation, marketing, and hazard maps.
However, the MAC address randomization systems introduced around 2014 make all conventional MAC-address-based crowd monitoring systems count the same device more than once. 
Therefore, there is a need to create a new crowd monitoring system tolerant to MAC address randomization to estimate the number of devices accurately. 
In this paper, Vision and TrueSight, two new crowd monitoring algorithms that estimate the number of devices, are proposed to prove that MAC-address-based crowd monitoring is still possible. In addition to probe requests, Vision uses data packets and beacon packets to mitigate the influence of randomization. Moreover, \textit{TrueSight} uses sequence numbers and hierarchical clustering to estimate the number of devices. The experimental results of this study show that even without installing any special software, Vision can gather 440 randomly generated MAC addresses into one group and count only once, and \textit{TrueSight} can estimate the number of devices with an accuracy of more than 75\% with an acceptable error range of 1.
	
\end{abstract}

\ccsdesc{Human and societal aspects of security and privacy}

\keywords{Crowd monitoring, Wi-Fi, MAC address}

\maketitle

\section{Introduction}
Wireless devices have become popular in recent times. According to research conducted by the Japanese government ~\cite{Japanese_government}, more than 80\% of people now have smartphones. Therefore, researchers are trying to analyze people’s positions and actions by gathering probe requests, a type of Wi-Fi packet sent by wireless devices ~\cite{Intro_a, Intro_b, Intro_c, Intro_d, Intro_e, Intro_f, Intro_g, Intro_h, Intro_i}. Probe requests are used to search for nearby access points (APs) and contain the devices’ media access control (MAC) addresses.

% どこかに入れた方がいい気がするが、どこにも入らない。削ろう。
% The format of Wi-Fi packets is shown as follows:

% \begin{itemize}
% 	\item Radiotap Header (18 octets)
% 	\item Frame Control (2 octets)
% 	\item Duration (2 octets)
% 	\item Destination Address (6 octets)
% 	\item Source Address (6 octets)
% 	\item BSSID (6 octets)
% 	\item Sequential Control (2 octets)
% 	\item Frame Body (variable length)
% \end{itemize}

The basic idea of the original crowd monitoring system was to use frame control to check whether a Wi-Fi packet was a probe request and gather MAC addresses in the source address. Researchers considered MAC addresses to be unique, and the number of people could be counted simply by counting the number of addresses. However, in 2014, vendors began putting MAC address randomization systems into their devices to protect users’ privacy.

For example, in 2015, Microsoft added a MAC address randomization system to Windows 10 ~\cite{Windows}. In Windows 10, random addresses are used not only for probe requests, but also for connecting to a network. To ensure that the client always uses the same address when connecting to a specific network, the following formula is used to calculate addresses:
	
$addr = SHA-256(SSID; macaddr; connId; secret)[:6]$
	
Here, \emph{SSID} is the name of the network, \emph{macaddr} is the original MAC address, and \emph{connId} is a parameter that changes if the user removes (and re-adds) the network from its preferred network list. \emph{secret} parameter is a 256-bits cryptographic random number generated during system initialization. This number is unique per interface and remains the same across reboots.

In 2014, there was a commit that added a MAC address randomization system into the Linux kernel ~\cite{Linux_a}. Furthermore, in 2015, wpa\_supplicant, a type of Wi-Fi management software for Linux, also started supporting randomization from version 2.4 ~\cite{Linux_b}.

In 2014, Apple added a MAC address randomization system to iOS 8 ~\cite{iOS_a}. In iOS 8, randomized addresses are only used when they are not connected to any AP and in sleep mode. In iOS 9, they are also used when not connected to any AP and in active mode ~\cite{iOS_b}. In iOS 14, iPadOS 14, and watchOS 7, Apple added per network MAC randomization support ~\cite{iOS_c, iOS_d}.

In 2015, Google added a MAC address randomization system to Android 6.0 ~\cite{Android_a}. In Android 6.0, randomized addresses are only used when they are not connected to any AP. Starting from Android 8.0, Android devices use randomized MAC addresses when probing for new networks and are not currently associated with a network. In Android 9, Google added per network MAC randomization support ~\cite{Android_b}.

When these systems are turned on, the MAC addresses inside probe requests change once every few minutes. This makes malicious users unable to trace devices using MAC addresses. However, it also makes all conventional MAC-address-based crowd monitoring systems count the same device more than once. 

Thus, this paper presents the design and implementation of a new crowd monitoring system that is tolerant to MAC address randomization. \textit{Vision} and \textit{TrueSight} were designed and implemented to show that it is still possible to develop a high-accuracy Wi-Fi packet-based crowd monitoring system.

The remainder of this paper is organized as follows.
Section~\ref{sec:related} presents an overview of the related works. 
Section~\ref{sec:issues_and_objectives} describes the goal of the study.
Section~\ref{sec:key_concept} introduces the key concept of the proposed method.
Section~\ref{sec:evaluation} shows how the method actually works.
Finally, Section~\ref{sec:conslusion} concludes the paper. 

\section{Related works}\label{sec:related}
Several studies have attempted to reduce the impact of MAC address randomization. Nakata et al. ~\cite{Senkou_a} proposed a method that uses frame control, MAC addresses, inter-frame arrival time, sequence numbers, and received signal strength indicator (RSSI) to estimate the number of devices.

Vanhoef et al. ~\cite{Senkou_b} clustered probe requests based on the fingerprints of their frame bodies and sequence numbers to estimate the number of devices. They classified probe requests by fingerprints first and then distinguished devices by sequence numbers and packet arrival times. However, if a device is connected to an AP, it will send both probe requests and data packets. This means that the sequence numbers in probe requests will have a considerable gap when connected, leading to low estimation accuracy.

Bernardos et al. ~\cite{Intro_i} studied the feasibility of MAC address randomization for associated devices in real-life conditions.
Freudiger ~\cite{Senkou_c} found that sequence numbers and timing information can be used to reidentify anonymized probes.
Guo et al. ~\cite{Senkou_d} used the inter-frame arrival time and sequence numbers to detect spoofing on a network with a semi-active method. Desmond et al. ~\cite{Senkou_e} fingerprinted devices using inter-frame time analysis alone.

% XXX,YYY,ZZZの関連研究をレビューしたが、本研究で目指す、AAAはこれまでになかった。みたいな文章が欲しい。

In this study, a new crowd monitoring method is proposed that uses data packets and beacons to separately consider devices that are already connected to APs and those not connected to APs. To the best of our knowledge, there are no studies considering this, and we believe that this will make a more accurate estimation system.

\section{Issues and objectives}\label{sec:issues_and_objectives}

In this study, the aim is to classify MAC addresses and estimate the number of devices accurately.

\subsection{MAC addresses Classification}

Many studies assume that all devices randomize their MAC addresses. However, when a device is connected to an AP, it can be ensured that its MAC address will not change randomly. This is because devices and APs use their MAC addresses to transfer Wi-Fi packets when connected. If they change their MAC addresses, the connection is terminated immediately, leading to a bad user experience. In this study, MAC addresses are classified into three groups.

\begin{itemize}
	\item Those of devices already connected to APs
	\item Those of devices not connected to APs
	\item Those of APs
\end{itemize}

To classify MAC addresses, beacon packets, data packets, and probe requests are used. They have the following properties, respectively:

\begin{description}
	\item [\textbf {Beacon packets}] are sent by APs. They are used to tell the surrounding devices their existence, and they contain APs' MAC addresses. Unlike devices, APs do not change their MAC addresses.
	\item [\textbf{Data packets}] are sent by both APs and devices. They are used to transfer data, and they contain both devices' MAC addresses and APs'. These packets are sent only when the connection is established.
	\item [\textbf{Probe requests}] are sent by devices. They are used to tell the surrounding APs their existence, and they contain the devices' MAC addresses.
\end{description}

The type of Wi-Fi packets can be distinguished by checking frame control.
It is necessary to have a system to observe whether a MAC address is randomly generated.

\subsection{Accurate estimation}

% 大幅な変更あり
% 他の論文に載っている方法よりも多くの種類のパケットを確認する必要があり、速度上では優位性がないと感じたため、fastを削除した。それに伴い、文を一部変更した。

The aim is to find a method to accurately estimate the number of devices. In the previous section, MAC addresses were classified into three groups. Because devices do not change their MAC addresses when connected to APs, the number of MAC addresses can be simply counted to obtain the number of connected devices. However, this method cannot be used for devices that are not connected to APs. Therefore, there is a need to find parameters other than MAC addresses to estimate the number. In this study, hard-to-modify parameters in the frame body and sequence numbers in sequential control are used to achieve this.

\section{Proposed Method}\label{sec:key_concept}

Our proposed method consists of two parts: \textit{Vision} and \textit{TrueSight}. \textit{Vision} classifies MAC addresses and observes frame bodies in probe requests to estimate the number of devices. \textit{TrueSight} observes sequence numbers in sequential control to count.

\subsection{Vision}

\begin{figure}[htbp]
	\centering
	\includegraphics[width=\linewidth]{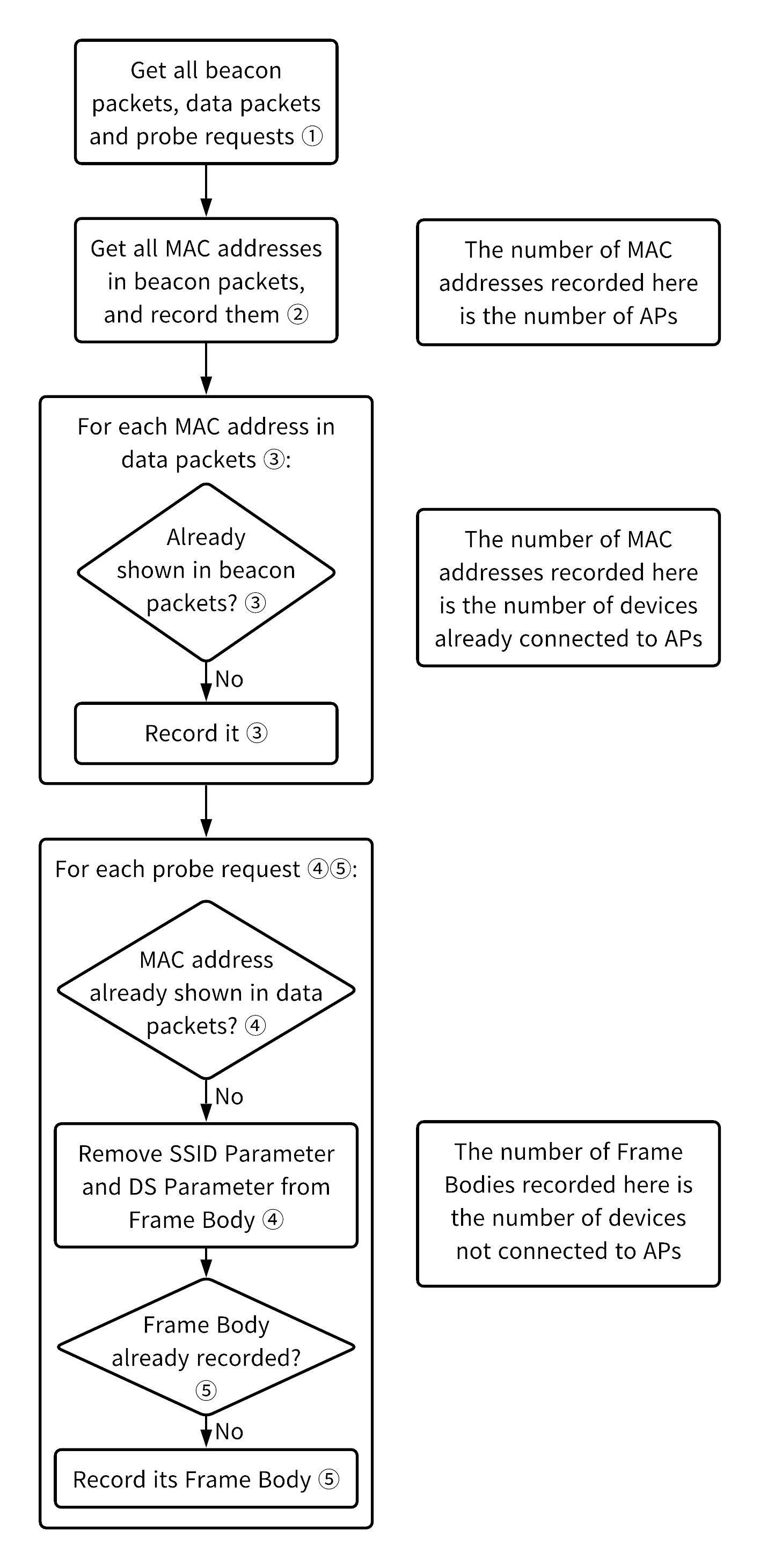}
	\caption{Flowchart of Vision}
	\label{fig:vision_flowchart}
\end{figure}

The essential idea behind \textit{Vision} is to classify the MAC addresses, as shown in Section 3.1. To do this, the following four elements in Wi-Fi packets must be checked: frame control, destination address, source address, and frame body. 
A flowchart of \textit{Vision} is shown in figure \ref{fig:vision_flowchart}.

\begin{figure}[htbp]
	\centering
	\includegraphics[width=\linewidth]{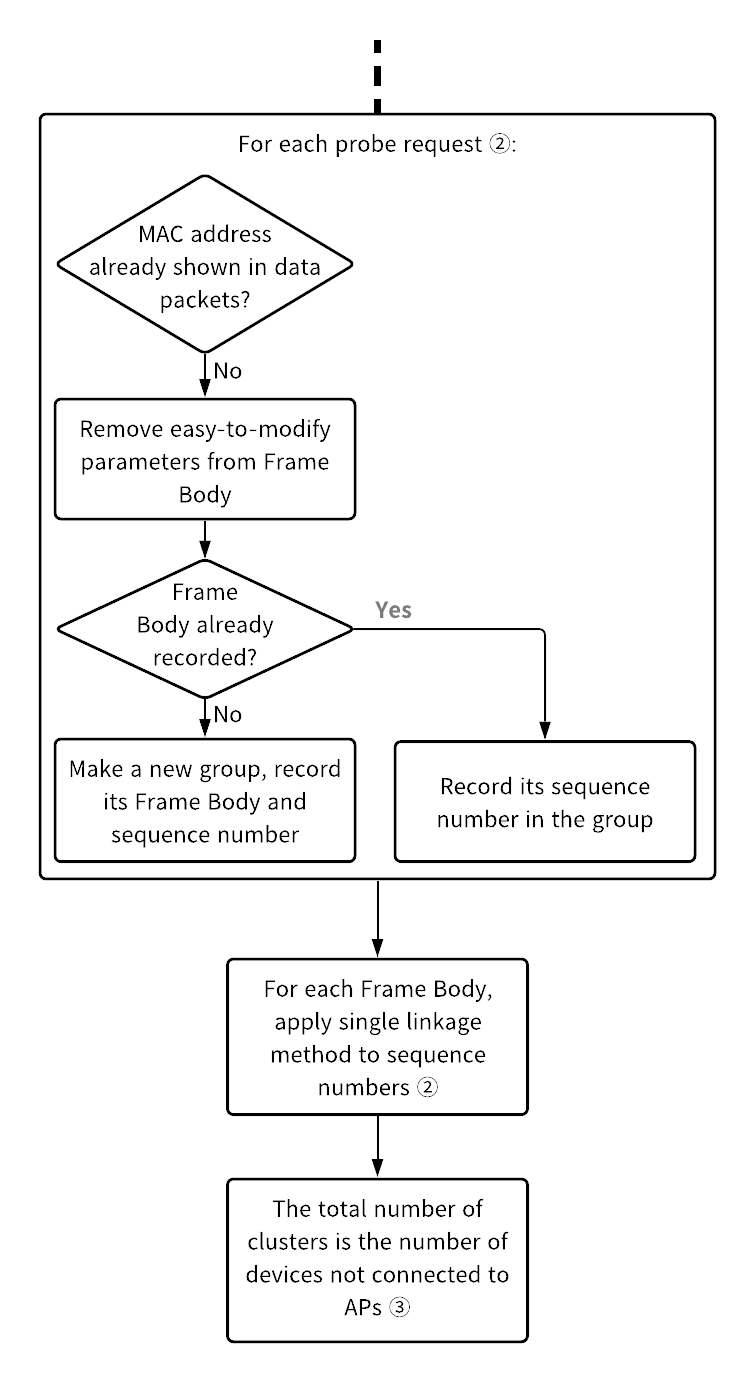}
	\caption{Flowchart of TrueSight}
	\label{fig:truesight_flowchart}
\end{figure}

The procedure of \textit{Vision} is shown as follows:

\begin{enumerate}
	\item Classify Wi-Fi packets into three groups, beacon packets, data packets, and probe requests, by checking frame control.
	\item Gather MAC addresses from the source address in all beacon packets. These are the MAC addresses of APs.
	\item Gather MAC addresses from the destination address and source address in all data packets and remove those of APs. The remaining MAC addresses belong to devices that have established connections with APs. The number of MAC addresses is the number of devices that have established connections with the APs.
	\item Gather MAC addresses from the source address in all probe requests and remove those of devices which have established connections with APs. The remaining MAC addresses belong to devices that have NOT established connections with the APs.
	\item Check the frame bodies of the probe requests sent by devices which have NOT established connections with APs. Remove the service set identifier (SSID) parameters and direct sequence spread spectrum (DSSS) parameters, and the number of the kinds of remainders will be the number of devices that have NOT established connections with APs.
	\item Calculate the sum of Step 3 and Step 5. The result is the number of devices nearby.
\end{enumerate}

In Step 5, it is assumed that the number of unique frame bodies is the number of devices that have NOT established connections with APs. This is because frame bodies in probe requests contain many characteristic and hard-to-modify parameters, and we can use them to distinguish devices.

% Table \ref{table:Frame_body_sample} shows major parameters in Frame Bodies in probe requests. For example, there is no problem modifying SSID Parameter, but if we modify Supported Rates, we might have a slower network connection and make users feel inconvenient. Since there are Tag number and Tag length to distinguish easy-to-modify parameters and hard-to-modify ones easily, Frame Body is an important and useful clue to count the number of devices.

% 調整あり

Table \ref{table:Frame_body_sample} shows the major parameters in frame bodies inside probe requests ~\cite{IEEE}. 
%Of all the parameters, we assume SSID parameter and DSSS parameter are uncharacteristic and easy to modify, remove them from frame bodies, and check the number of the kinds of remainders to estimate the number of devices accurately.
Of all the parameters, we remove the SSID parameter and DSSS parameter from frame bodies because we assume they are easy to modify and do not have characteristics for distinguishing between devices. 
Then, \textit{Vision} check the number of the kinds of remainders to estimate the number of devices accurately.

\begin{table*}[t]
	\begin{center}
		\caption{Major parameters in Frame Body inside probe requests}
		\label{table:Frame_body_sample}
		\tabcolsep = 1mm
		\begin{tabular}{c|c|c|c}
			\hline
			\begin{tabular}{c} Tag number \\ (1 octet) \end{tabular}  & (Tag name) & \begin{tabular}{c} Tag length \\ (1 octet) \end{tabular}  & \begin{tabular}{c} Contents \\ (Length is specified in tag length) \end{tabular} \\ \hline \hline
			0                             & \begin{tabular}{c} Service Set Identifier \\ (SSID) Parameter \end{tabular}           & Less than 256 & SSID that the device wants to connect with                           \\ \hline
			1                             & Supported Rates          & 8 or less     & Data transfer rates which the device supports                         \\ \hline
			3                             & \begin{tabular}{c} Direct Sequence Spread Spectrum \\ (DSSS) Parameter \end{tabular}  & 1             & \begin{tabular}{c}Device's channel setting                            \\ when sending the probe request \end{tabular}\\ \hline
			50                            & Extended Supported Rates & Less than 256 & \begin{tabular}{c}If a device supports eight or more data transfer rates, \\ this parameter will be used\end{tabular}\\ \hline
			221                           & Vendor Specific          & Less than 256 & \begin{tabular}{c}Vendor specific information                         \\ (e.g. Who made this device)\end{tabular}\\ \hline
		\end{tabular}
	\end{center}
\end{table*}

\subsection{TrueSight}

Because devices with the same model have a high chance of using the same frame body, \textit{Vision} cannot distinguish them properly in Step 5. Thus, another parameter must be found to distinguish them.

In \textit{TrueSight}, another element is used in Wi-Fi packets: sequential control. It contains sequence numbers that show how many Wi-Fi packets a device has sent. If a device is not connected to any AP, it will not send any packet other than probe requests, so there is a high probability that its sequence numbers will be consecutive. For two reasons, the single linkage method, a type of hierarchical clustering method can be used to estimate the number of devices accurately.

\begin{itemize}
	\item Sequence numbers in probe requests sent by the same device have a high chance to be consecutive or close values.
	\item Sequence numbers in probe requests sent by different devices have little chance to be close values
\end{itemize}

Flowchart of \textit{TrueSight} is shown in figure \ref{fig:truesight_flowchart}.

The procedure of \textit{TrueSight} is shown as follows:

\begin{enumerate}
	\item Follow the procedure of Vision until Step 5.
	\item For each frame body, gather sequence numbers from sequential control in probe requests, set a proper threshold and apply single linkage method to them. Then, the number of clusters is recorded.
	\item Calculate the total number of clusters. This is the number of devices that have NOT established connections with APs.
	\item Calculate the sum of Step 5 in Vision and Step 3 in TrueSight. The result is the number of devices nearby.
\end{enumerate}

\section{Evaluation}\label{sec:evaluation}

\subsection{Reducing the impact from MAC address randomization}

To observe whether \textit{Vision} can reduce the impact of MAC address randomization and estimate the number of devices using data packets, beacon packets, and probe requests, a system was created to capture and analyze packets and check the difference between the number of MAC addresses not connected to APs and the number of frame bodies.

\begin{figure}[htbp]
	\centering
	\includegraphics[width=\linewidth]{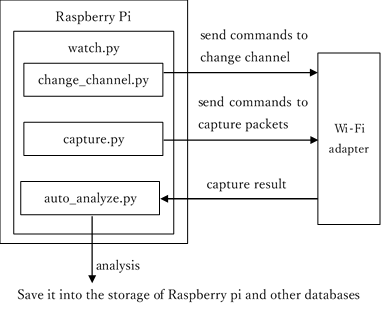}
	\caption{System made for the experiment}
	\label{fig:vision_system}
\end{figure}

We experimented to check if \textit{Vision} can estimate the number of devices properly in the laboratory, from September 15, 2020, to September 21, 2020. Figure~\ref{fig:vision_system} shows the system used in the experiment. This system consists of two devices, a Raspberry Pi that can execute \textit{Vision}, and a Wi-Fi adapter that can turn on the monitor mode and capture packets.

In the part of the Raspberry Pi, we installed \textit{Raspberry Pi OS} as operating system, \textit{aircrack-ng} and \textit{iwconfig} for configuring the Wi-Fi adapter, \textit{tshark} and \textit{tcpdump} for capturing Wi-Fi packets, and \textit{Python 3.7} for making \textit{Vision}. To make \textit{Vision}, we used \textit{datetime}, \textit{json}, \textit{os}, \textit{requests}, \textit{subprocess}, \textit{sys}, \textit{time}, \textit{watchdog} Python libraries.

For the Wi-Fi adapter, we used the Panda 300Mbps Wireless N USB Adapter Model PAU05. It has a Ralink RT5372 chipset and can capture 802.11b/g/n (2.4GHz) Wi-Fi packets.

% \subsubsection{Equipment}

% In this experiment, we used the following two devices.

% \begin{itemize}
% 	\item Raspberry Pi 3 Model B
% 	      \begin{itemize}
% 	      	\item OS: Raspberry Pi OS
% 	      	\item Used packages: aircrack-ng, iwconfig, tshark, tcpdump, Python 3.7
% 	      	\item Used Python libraries: datetime, json, os, requests, subprocess, sys, time, watchdog
% 	      \end{itemize}
% 	\item Panda 300Mbps Wireless N USB Adapter Model PAU05 (Used to capture packets)
% 	      \begin{itemize}
% 	      	\item Chipset: Ralink RT5372
% 	      	\item Able to capture 802.11b/g/n (2.4GHz) Wi-Fi packets
% 	      \end{itemize}
% \end{itemize}

% \subsubsection{Programs}

% In this experiment, we used the following four Python programs.

% \begin{itemize}
% 	\item watch.py
% 	      \begin{itemize}
% 	      	\item turn on monitor mode for Wi-Fi adapters
% 	      	\item manage three other Python programs
% 	      	\item send pcap files to analyze.py
% 	      \end{itemize}
% 	\item change\_channel.py
% 	      \begin{itemize}
% 	      	\item change the monitoring channel of the Wi-Fi adapter every 20 seconds
% 	      \end{itemize}
% 	\item capture.py
% 	      \begin{itemize}
% 	      	\item capture probe requests, data packets and beacon packets and save them as pcap files every 10 minutes
% 	      \end{itemize}
% 	\item auto\_analyze.py
% 	      \begin{itemize}
% 	      	\item get pcap files from watch.py, analyze and export analysis results
% 	      \end{itemize}
% \end{itemize}

\subsubsection{Experiment result}

\begin{figure}[h]
	\centering
	\includegraphics[width=\linewidth]{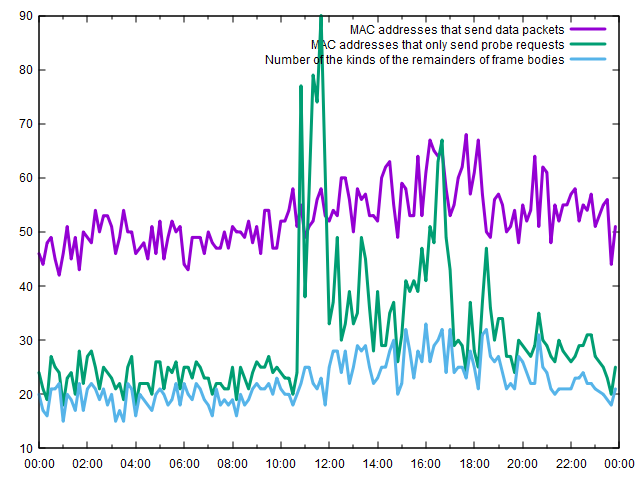}
	\caption{Analysis result in Sep. 19th, 2020}
	\label{fig:vision_result}
\end{figure}

Of all the data, the analysis data on September 19, 2020 is shown in figure \ref{fig:vision_result}, which clearly shows the influence of MAC address randomization. The influence is also detected on other days, but the data on September 19, 2020 is the most obvious one. 
From figure \ref{fig:vision_result}, it can be seen that there is not much difference between the number of MAC addresses that only send probe requests and the number of unique frame bodies, until 10 A.M. However, they differ significantly after that time. By checking every packet detected, it was found that there were 440 MAC addresses that sent packets with the same frame body from 10:40 to 17:10. Since all of our lab's devices were connected to Wi-Fi routers during the experiment, there is no doubt that these packets were sent by outsiders. All 440 MAC addresses were detected only once, and it was concluded that these MAC addresses were randomly generated by one device, and \textit{Vision} succeeded in reducing the impact of MAC address randomization.

\subsection{Number of devices estimation}

% 何を検証することを目的にどんな実験をするか、書く。Done

% To observe how accurately \textit{TrueSight} can estimate the number of devices using sequence numbers, a simulator was implemented to generate datasets. We let \textit{TrueSight} analyze them, and checked how many times \textit{TrueSight} estimated the number accurately.
In this section, we evaluate how accurately \textit{TrueSight} can estimate the number of devices using sequence numbers. 
Because it is challenging to prepare many devices using the same model, we implemented a simulator to generate a realistic dataset with many nodes. Then, we verified how many times \textit{TrueSight} estimated the number accurately with the dataset.

\subsubsection{Dataset preparation}

Our simulator generates the dataset as follows:

\begin{enumerate}
	\item Decide the number of the devices (X) and the probe requests sent per device (Y).
	\item Choose a random integer from 0 to 4095 and record it. Because sequence numbers are within the range of 0 to 4095, we cannot choose a number out of its range.
	\item Add a random integer from 1 to 49 to the last number recorded and record the result of addition. If the new number is 4096 or more, subtract 4096 from it first, and then record the result of subtraction. This is because in the experiment of Vision, there was no device that the maximum gap between any pair of adjacent sequence numbers was 50 or more.
	\item Repeat Step 3 for (Y-1) times, respectively. This step aims to imitate a situation in which a device sends Y probe requests.
	\item Repeat from Step 2 to Step 4 for (X-1) times. This step aims to imitate the situation in which X devices send probe requests.
	\item For all the numbers recorded, set the threshold to 50 and apply single linkage method, one kind of hierarchical clustering methods.
	\item The number of the clusters made by Step 6 is the number of devices. This is because it is guaranteed in Step 3 that the maximum gap between any pair of adjacent sequence numbers sent by the same device is less than 50.
\end{enumerate}

\begin{table*}[htbp]
	\begin{center}
		\captionsetup{justification=centering}
		\caption[Experiment result when fixing the value of X to 5 and changing the value of Y from 1 to 6]{Experiment result when fixing the value of X to 5 and changing the value of Y from 1 to 6}
		\label{additional_experiment_1}
		\tabcolsep = 1mm
		\begin{tabular}{c|c|c|c|c|c|c|c|c}
			\multirow{3}{*}{\begin{tabular}[c]{@{}c@{}}The \\ number \\ of devices \\ (X)\end{tabular}} & \multirow{3}{*}{\begin{tabular}[c]{@{}c@{}}Probe \\ requests \\ per device \\ (Y)\end{tabular}} & \multicolumn{6}{c|}{Estimation result (x)}                                                                                           & \multirow{3}{*}{\begin{tabular}[c]{@{}c@{}}Total \\ number \\ of trials \\ (N)\end{tabular}} \\ \cline{3-8}
			                   &   & 1              & 2           & 3     & 4     & 5              & 6   &        \\ \cline{3-8}
			&                                                                                          & \multicolumn{6}{c|}{\begin{tabular}[c]{@{}c@{}}The number of times that \\ TrueSight estimates there are x devices (n)\end{tabular}} &                                                                                        \\ \hline
			\multirow{6}{*}{5} & 1 & \ \ \ 0\ \ \ \  & \ \ 4\ \ \  & 152 & 2,124 & \textbf{7,720} & 0 & 10,000 \\ \cline{2-9} 
			                   & 2 & \ \ \ 0\ \ \ \  & \ \ 13\ \ \  & 332 & 2,785 & \textbf{6,666} & 204 & 10,000 \\ \cline{2-9} 
			                   & 3 & \ \ \ 2\ \ \ \  & \ \ 29\ \ \  & 546 & 3,246 & \textbf{5,824} & 353 & 10,000 \\ \cline{2-9} 
			                   & 4 & \ \ \ 1\ \ \ \  & \ \ 66\ \ \  & 785 & 3,704 & \textbf{4,986} & 458 & 10,000 \\ \cline{2-9} 
			                   & 5 & \ \ \ 2\ \ \ \  & \ \ 102\ \ \  & 1,066 & 3,922 & \textbf{4,352} & 556 & 10,000 \\ \cline{2-9} 
			                   & 6 & \ \ \ 4\ \ \ \  & \ \ 141\ \ \  & 1,334 & 4,089 & \textbf{3,857} & 575 & 10,000 \\
		\end{tabular}
	\end{center}
\end{table*}

\begin{table*}[htbp]
	\begin{center}
		\captionsetup{justification=centering}
		\caption[Experiment result when fixing the value of Y to 2 and changing the value of X from 1 to 8]{Experiment result when fixing the value of Y to 2 and changing the value of X from 1 to 8}
		\label{additional_experiment_2}
		\tabcolsep = 1mm
		\begin{tabular}{c|c|c|c|c|c|c|c|c|c|c|c}
			
			\multirow{3}{*}{\begin{tabular}[c]{@{}c@{}}The \\ number \\ of devices \\ (X)\end{tabular}} & \multirow{3}{*}{\begin{tabular}[c]{@{}c@{}}Probe \\ requests \\ per device \\ (Y)\end{tabular}} & \multicolumn{9}{c|}{Estimation result (x)}                                                                                                  & \multirow{3}{*}{\begin{tabular}[c]{@{}c@{}}Total \\ number \\ of trials \\ (N)\end{tabular}} \\ \cline{3-11}
			  &                    & 1              & 2              & 3              & 4              & 5              & 6              & 7              & 8              & 9   &        \\ \cline{3-11}
			&                                                                                          & \multicolumn{9}{c|}{\begin{tabular}[c]{@{}c@{}}The number of times that \\ TrueSight estimates there are x devices (n)\end{tabular}}      &                                                                                        \\ \hline
			1 & \multirow{8}{*}{2} & \textbf{9,940} & 60             & -              & -              & -              & -              & -              & -              & -   & 10,000 \\ \cline{1-1} \cline{3-12} 
			2 &                    & 358            & \textbf{9,523} & 119            & -              & -              & -              & -              & -              & -   & 10,000 \\ \cline{1-1} \cline{3-12} 
			3 &                    & 25             & 1,023          & \textbf{8,786} & 166            & -              & -              & -              & -              & -   & 10,000 \\ \cline{1-1} \cline{3-12} 
			4 &                    & 2              & 121            & 1,912          & \textbf{7,781} & 184            & -              & -              & -              & -   & 10,000 \\ \cline{1-1} \cline{3-12} 
			5 &                    & 0              & 13             & 334            & 2,771          & \textbf{6,680} & 202            & -              & -              & -   & 10,000 \\ \cline{1-1} \cline{3-12} 
			6 &                    & 0              & 4              & 59             & 766            & 3,500          & \textbf{5,470} & 201            & -              & -   & 10,000 \\ \cline{1-1} \cline{3-12} 
			7 &                    & 0              & 0              & 9              & 183            & 1,312          & 3,961          & \textbf{4,337} & 198            & -   & 10,000 \\ \cline{1-1} \cline{3-12} 
			8 &                    & 0              & 0              & 1              & 52             & 457            & 1,957          & 4,115          & \textbf{3,247} & 171 & 10,000 \\
		\end{tabular}
	\end{center}
\end{table*}

Clusters made in Step 6 guarantees that the minimum number is 1, and the maximum number is $min([\frac{4096}{50}]+1, X+1)$. For the minimum case, this will occur when all sequence numbers are so close that any pair of adjacent sequence numbers differs by less than 50. For the maximum case, there are two possibilities.

\begin{itemize}
	\item  Any pair of adjacent sequence numbers differs by exact 50. In this case, the number of devices is $[\frac{4096}{50}]+1 = 81+1 = 82$.
	\item  Any pair of sequence numbers sent by different devices differs by more than 50, and one of the devices reached 4095 and went back to 0. In this case, this device is counted twice, and the number of devices will be $X+1$.
\end{itemize}

To reduce the influence of random number bias, this experiment was carried out 10,000 times, and the number of times \textit{TrueSight} estimated the number of devices correctly was recorded.

We used \textit{Python 3.7} to implement the simulator, which is made up of \textit{numpy}, \textit{pandas}, \textit{scipy}, \textit{matplotlib}, \textit{random}, and \textit{collections} Python libraries.

% \subsubsection{Equipment}

% In this experiment, we used the following device.

% \begin{itemize}
% 	\item Raspberry Pi 3 Model B
% 	      \begin{itemize}
% 	      	\item OS: Raspberry Pi OS
% 	      	\item Used package: Python 3.7
% 	      	\item Used Python libraries: numpy, pandas, scipy, matplotlib, random, collections
% 	      \end{itemize}
% \end{itemize}

\subsubsection{Experiment result}

Table \ref{additional_experiment_1} is the experiment result when fixing the value of X and changing the value of Y. Bold text indicates that the estimated number of devices matches with the actual one. From the result, we can say that when the number of probe requests per device goes up, the accuracy goes down. This is because the more probe requests are sent, the higher is the probability of sequence numbers becoming close. Nevertheless, \textit{TrueSight} can estimate the number of devices to a certain extent. If an acceptable error range of 1 is set, the accuracy will be over 80\%. To make the estimation more accurate, probe requests must be captured not only at least once, but also as few as possible. According to the research by Julien Freudiger ~\cite{Senkou_d}, devices send their probe requests about once a minute. Therefore, to use \textit{TrueSight} in a practical situation, setting capture time to one minute is sufficient.

Table \ref{additional_experiment_2} is the experiment result when fixing the value of Y and changing the value of X. From the result, we can say that when the number of devices goes up, the accuracy goes down. This is because the more devices there are, the higher the chance of sequence numbers becoming close. Nevertheless, \textit{TrueSight} can estimate the number of devices to a certain extent. If we set an acceptable error range of 1, the accuracy will be more than 75\%.

From the two experimental results, we can say that \textit{TrueSight} is able to estimate the number of devices. To make the estimation more accurate, we need to capture probe requests not only at least once, but also as few as possible.

\section{Conclusion and Future works}\label{sec:conslusion}

In this study, a new crowd monitoring system was developed that is tolerant to MAC address randomization. It was found that using characteristic and hard-to-modify parameters inside probe requests is a good way to reduce the impact of randomization.
Furthermore, our system only counts the number of devices and does not identify them. This means that if a person turns on the randomization system, our system can count their device once, but it cannot identify them because devices with the same model have a high chance of using the same frame body. Therefore, our system enables users’ privacy and crowd monitoring to coexist.
Moreover, our system does not require users to install special software. It only requires users to turn on the Wi-Fi function to gather information. This means that more information can be obtained compared to other systems.
We believe \textit{Vision} and \textit{TrueSight} open up new directions of research on Wi-Fi-packet-based crowd monitoring system.

% Our future works includes XXX

Our future work will include checking the difference between the estimated number of devices by Vision and the actual one, and testing TrueSight in practical situations. It was verified that Vision could estimate the number of devices while reducing the impact of MAC address randomization. However, the number of devices was not counted, so it could not be checked whether the estimated number was accurate. In addition, it was verified that TrueSight could distinguish devices using the same model, but it was only tested in the simulated data. Therefore, there is a need to test this method in actual situations to determine whether it works well.

\end{document}